\documentclass{evn2004}
\usepackage{graphicx}
\begin{document}
\title{Polarisation in CSS/GPS radio sources}

\author{D. Dallacasa\inst{1,2}
}
\institute{
Dipartimento di Astronomia, Universit\`a di Bologna, via Ranzani 1,
I-40127, Bologna, Italy \and 
Ist. di Radioastronomia -- CNR, via Gobetti 101, I-40129, Bologna,
Italy 
}

\abstract{
The polarisation properties of the intrinsically small ($<$ 20~kpc)
radio sources are briefly discussed. At centimetric and decimetric
wavelengths, substantial polarised emission is often detected in
quasars and, at a lower level, in radio sources identified with
galaxies with projected linear sizes in excess of a few kpc. The
projected linear size at which some polarised emission is detected
decreases with observing frequency. \\ 
High Rotation Measures are commonly found in these sources, implying
high electron/ion densities and/or strong interstellar magnetic
fields. For the objects found unpolarized at resolutions of the order
of a few pc, either the source magnetic field is highly disordered or
the scale of ambient magnetic field cells is not resolved.

}
\titlerunning{Polarisation in CSS/GPS radio sources}
\maketitle
\section{Introduction}
\subsection{On CSS/GPS radio sources}
Compact Steep-Spectrum (CSS) and GHz Peaked-Spectrum (GPS) radio  
sources are powerful (P$_{1.4 {\rm GHz}}$$>$10$^{25}$ W/Hz)
extragalactic objects characterized by a small physical size
(typically $\leq$20 $h^{-1}$ kpc\footnote{H$_{0}=$ 100$h$ km s$^{-1}$
Mpc$^{-1}$ and $q_{0}$ = 0.5 have been assumed.} and  $\leq$1 $h^{-1}$
kpc, respectively for CSSs and GPSs). Their optically thin radio
emission has a steep spectrum ($\alpha > 0.5$, $S\sim\nu^{-\alpha}$)
and turns over around hundreds of MHz (CSSs) or around 1 GHz
(GPSs). \\ 
They represent a significant fraction ($\simeq$ 15--30\%) of the
radio source population at cm wavelengths and are generally associated
with distant objects ($z>0.2$), both galaxies and quasars. The
fraction of quasars increases with flux density and peak frequency,
and reaches about 50\% in bright CSS and GPS samples (see
below). Given that quasars are generally found at a higher redshift than
galaxies, the intrinsic peak frequency is higher in quasars
than in galaxies (see O'Dea \cite{odea}).\\ 
It is believed that at least those with two-sided radio emission
represent the young stage in the radio source evolution, from Compact
(size $<$1 $h^{-1}$ kpc) to Large (size $>$20 $h^{-1}$ kpc) Symmetric
Objects (Fanti et al. \cite{cf95}; Readhead et al. \cite{read};
Snellen et al. \cite{sne00,sne03}).     
The measurements of hot-spot proper motions in Compact Symmetric
Objects (CSO) characterised by projected linear sizes of the order of 
a few hundreds pc (Polatidis \& Conway \cite{pol}) and with
separation velocities of the outer edges of $\sim$0.1--0.4~$h^{-1}$~c,
imply kinematic ages of the order of than 3$\times$10$^{3}$ years or
even less for the smallest objects. 
Among the sources considered by Polatidis \& Conway, ten have proper
motion detected in excess of 0.1c, while three  have upper limits to
the separation speed between 0.05 and 0.1c.
\\
On the other hand, the spectral ages, based on synchrotron ageing
arguments and pinpointed by a break in the total radio spectrum, are
of the order of $10^{3}-10^{5}$~years for the sources studied in 
Murgia et al. (\cite{mmu}) and Murgia (\cite{mm}), in very good
agreement with kinematic ages derived for the objects in common with
Polatidis \& Conway (\cite{pol}). All these findings support the
``youth'' scenario, against the idea of ``frustration'' (e.g. Baum et
al. \cite{ba90}), requiring a particularly dense ambient medium capable of
preventing the source growth and then confining the radio emission to
subgalactic size for the its whole lifetime. 

\subsection{On polarisation properties}

Synchrotron radiation is intrinsically {\bf linearly} polarised since
each electron is forced to oscillate on a plane perpendicular to the
local magnetic field lines.  
In general, if linearly polarised radiation crosses an external region
where magnetic field and free electrons coexist, the orientation of
the intrinsic polarisation vector is rotated by:
$$ \Delta\theta \sim {\rm RM} \times \lambda^2 ~~~({\rm rad})$$
where RM stands for Rotation Measure:
$$ {\rm RM} = 812 \int_l n_eB_{||} {\rm dl}
~~~({\rm rad/m^2})$$
where $B_{||}$ is the field component along the line of sight.
This effect takes place inside the radio source itself, or may
happen along the path traveled by the polarised radiation. \\
An ensemble of electrons in a perfectly ordered magnetic field 
may emit radiation linearly polarised up to 70\% since we have to
consider that the photons emitted at distances larger than the surface
of the volume filled in by the radio source itself must travel in a
region where the source magnetic field and the ions (electrons) rotate
the polarisation plane. The presence of hot (i.e. ionised) thermal
plasma within the volume of the radio source may depolarise the
radiation in a manner extremely difficult to predict. \\

On the contrary, if a Faraday screen is located along the line of
sight, the behaviour of the polarised emission strongly depends on the
geometry and on the characteristic scale of the magnetic field
fluctuations. If the cells (i.e. regions where the field can be
considered uniform) of this Faraday screen are not resolved by
the observation the observed fractional polarisation can be reduced to
a much lower level and even down to zero, given that differential
rotation within the beam may produce a null average of the
polarisation vectors, one for each cell (beam depolarisation). 

Otherwise, a partially/fully disordered magnetic field within the
radio source may largely reduce the fractional polarisation, and even
completely depolarise the radiation leaving the radio source.

Fractional polarisation close to the theoretical maximum (70\%) have been
found in relaxed regions of extended radio sources, i.e in the radio
lobes and backflow tails, on scales of tens of kpc. On smaller scales,
relevant to CSS/GPS radio sources, instead, such values have never
been observed.

In general, bright CSS/GPS sources are known to possess either little
or no linearly polarised emission, with a few outstanding exceptions
(quasar only) which will be briefly discussed further down. \\
No circular polarisation (CP) has ever been detected at a fraction of
a \% level in CSS/GPS sources. Indeed CP is detected in the cores or
at the jet base in some blazars, i.e. sources where relativistic
beaming plays a major role. It is well known instead that such effect
is negligible in CSS/GPS sources.   

Given their small linear size, it is clear that the interstellar
medium of the host galaxy acts as main Faraday screen on the linearly
polarised emission, since the dominant contribution to the intervening
column density generally comes from the radio source host. As
previously stated, if the characteristic scale of the
``inhomogeneities'' in this medium is not resolved by the 
observations then the beam depolarisation can reduce the degree of
polarisation, even down to zero.

Further, since the jets of the radio source are still digging their
way within the host galaxy, they could induce shocks into the ISM 
producing a cocoon of ionised material surrounding the working surface
of the radio source (Bicknell et al. \cite{bdo}). This is more relevant
within the NLR where the density is larger and there are plenty of
clouds of ionised material.

VLA observations at cm wavelengths have demonstrated that some CSS
sources posses high integrated RM (Taylor et al. \cite{tit92}), in the 
range of the thousands rad/m$^2$. The only other radio sources with
comparably high RM values are found at the center of rich clusters of
galaxies where high electron/ion densities (the very hot intergalactic
gas responsible for the diffuse X-ray emission) and intergalactic
magnetic fields around 10~$\mu$G can be found. The typical tangling
scale of this screen is between 1-100 kpc.\\
It is expected to be much
smaller in the region surrounding CSS/GPS radio sources, and this, in
turn, would imply a substantial beam depolarisation. High frequency
(tens of GHz) polarisation observations with pc-scale resolution are
therefore a very important tool to study the linear polarisation in
intrinsically small sources.

Lobe asymmetries coupled to polarisation information may be of great
importance in the study of the role played by jet/ambient
interaction, since the (more) polarised side is expected to trace
the approaching side of the radio source, which is expected to be
slightly brighter and larger (and at a greater distance to the core)
with respect to the receding side, in standard source models.

\section{Samples of CSS/GPS radio sources}
The historical samples of intrinsically small radio sources have
selected the brightest objects of this class (the CSSs in Fanti et
al. \cite{f90}; the GPSs in Stanghellini et al. \cite{cstan}; we can
also include the High Frequency Peakers - HFPs - in Dallacasa et
al. \cite{hfps}, namely radio sources with spectra peaking at
frequencies higher than a few GHz). In these bright
samples the fraction of sources identified with quasars is around 50\%
with a tendency to increase with peak frequency. 
Going to lower flux densities the quasar fraction decreases
progressively (Snellen et al. \cite{snsn}; Marecki et al. \cite{mar};
Fanti et al. \cite{fanti01}) indication of a less extreme population.

If we refer to the Unified Scheme model (e.g. Urry and Padovani
\cite{up95}), CSS/GPS galaxies have the jet axis close to the plane of
the sky, while the CSS/GPS quasars are at intermediate angles; small
sources with the jet axis close to the line of sight would appear as
flat spectrum radio quasars. Therefore it is expected that the projected
linear size in radio quasars is somehow shortened by projection
effects, although the radio steep spectrum and a general stability of
the flux density ensure that relativistic beaming have marginal
effects in the vast majority of the CSS quasars. 

Some support to this picture comes from the radio morphologies of
CSS/GPS radio sources, observed at kpc resolution with the VLA (and
MERLIN) and at pc resolution with VLBI. 
The radio emission in galaxies is dominated by mini-lobes. They often
hosts more compact structures which can be considered as hot-spots
given that their local spectrum is well fitted by a power-law (Orienti
et al. \cite{mo04}; Murgia et al. in preparation). Cores are generally
weak and often remain undetected. In a few cases also jets are
visible.
The radio emission in CSS/GPS quasars is instead characterised either
by jets as a whole or by individual knots. These jets sometimes show
bends and contribute to most of the total source luminosity. Cores are
detected more frequently than in CSS/GPS galaxies and have a larger
contribution (but still at a few \% level) to the source total flux
density, although there are very few rare cases where the core is 
the most luminous component.
These morphological properties are in agreement with the expectations
from the Unified Scheme models.

Studies on the forementioned samples have led to build the following
paradigm to describe the polarisation properties of CSS/GPS sources:

\begin{enumerate}
\item Galaxies are generally unpolarized or, at least, strongly
  depolarised 
\item Quasars are either highly polarised (up to $\sim$10\%) or
completely depolarised
\item High RM (exceeding 1000 rad/m$^2$) are generally found in
  CSS/GPS sources 
\item Polarisation and flux density variability are not common in
  CSS/GPS radio sources
\end{enumerate}

\section{Integrated linear polarisation}
\begin{figure*} 
\centering
\includegraphics[height=10cm]{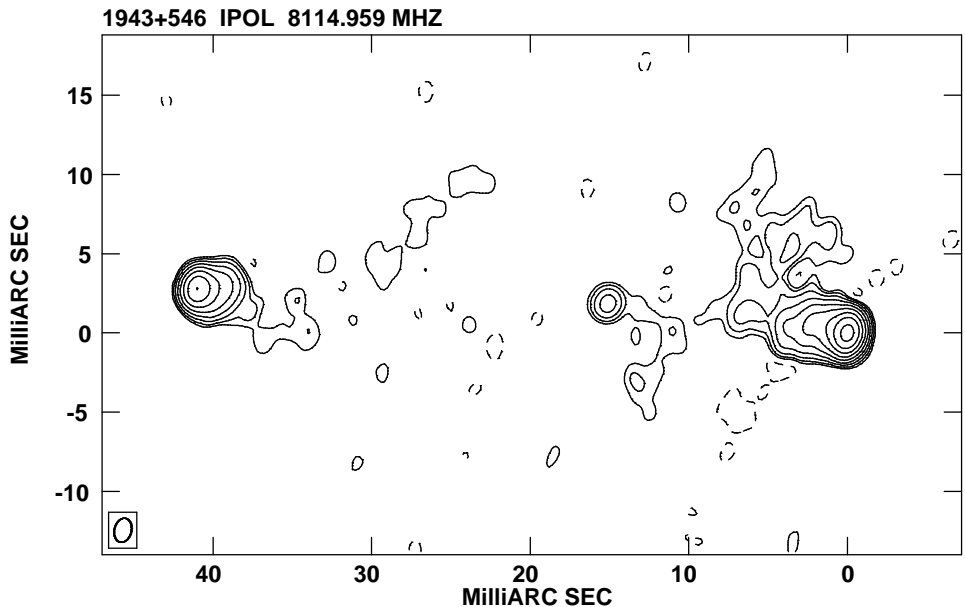}
\caption{Full resolution total intensity image of
  \object{B1943+546}. The first contour is $\pm$0.7 mJy/beam, and the
  levels scale up by a factor of 2. Some more diffuse emission from
  the lobes is resolved out}  
\label{IPOL}
\end{figure*}

One of the characteristics of CSS/GPS galaxies is the very low or even
absent linear polarisation at cm wavelengths. This property can be
related to the source growth model and then relates the observed
polarisation to the projected linear size, that is, in turn, related
to the turnover frequency (see e.g. O'Dea \cite{odea}).
Based on the work of Stanghellini et al. (\cite{cstan}), it is clear
that GPS galaxies have upper limits of 0.3\%  (but many as low as
0.1\%) to the integrated fractional polarisation at 4.9 and 8.5
GHz. The only object (\object{B1323+321}) detected in their
polarimetric VLA observations also belong to the CSS sample of Fanti
et al. (\cite{f90}). GPS quasars are generally weakly polarised,
around 1\% level, and about one third is completely unpolarised.
A systematic work on the polarisation of CSS sources of Fanti et
al. (\cite{f90}) has not been carried out. We can refer to van Breugel
et al. (\cite{vb84};\cite{vb92}) and L\"udke et al. (\cite{el}) to
find out that CSS galaxies are weakly polarised if not at all; CSS
quasars instead have outstanding sources for which cases of percentage of
polarisation as high as 10\% are observed even at decimetric
wavelengths. A few examples will be discussed further down.
A number of these CSS quasars are used as calibration sources for both
flux density and absolute orientation of the electric vector of the
linear polarisation.
A requirement for such use is to have a {\bf stable} flux density and
polarisation (in both intensity and orientation of the electric
vector), which can be considered a quite unusual characteristic if
compared to what is seen flat spectrum radio quasars and BL Lacs, in
which both quantities may vary substantially (Cawthorne et
al. \cite{caw93}). This can be considered a further evidence for the
different nature of CSS quasars, in which cores and components at the
jet base little contribute at the total source flux density and
polarisation,.

An interesting view has been given by Cotton et al. (\cite{wdc}) and
Fanti et al. (\cite{cf04}) who studied the B3-VLA CSS sources sample
(Fanti et al. \cite{fanti01}). These are sources with radio luminosities generally lower
than in the Fanti et al. (\cite{f90}), and the fraction of galaxies
(or empty fields) is much larger. By using the NVSS data, Cotton et
al. (\cite{wdc}) found that at 20 cm all sources with projected linear
size smaller than 6 kpc are not polarised; part of the larger sources
instead show some amount of polarisation. The analysis of VLA
A-configuration data presented by Fanti et al. (\cite{cf04}) at 6 and
3.6 cm (see their Fig. 8) demonstrates that the cutoff, quite sharp at
20 cm, moves inward the radio source and becomes shallower at
progressively shorter wavelengths. In particular, at 3.6 
cm the median level of polarisation is 4\% for the sources with
projected linear size exceeding 5 kpc. For many of these sources with
substantial polarisation even in the NVSS, it is possible to study the
RM: only about 50\% of the sources follow the expected $\lambda^2$
law, and this might be due to a complex field geometry and/or
differential effects within the radio source. Only about 20\% of the
sources have intrinsic rotation measure exceeding 1000 rad/m$^2$, and
in these cases the depolarisation gets substantial at lower
frequencies (Fanti et al. \cite{cf04}). 

\section{Parsec scale polarisation}

A possibility to attribute polarised emission to the smallest CSS/GPS
sources despite their low or absent integrated linear polarisation is
to invoke beam depolarisation.\\
A viable method to investigate this possibility consists of very high
resolution observations capable to resolve Faraday screen with scales
of the order of a few parsec. This implies the use of VLBI.

\begin{figure*} 
\centering
\includegraphics[height=10cm]{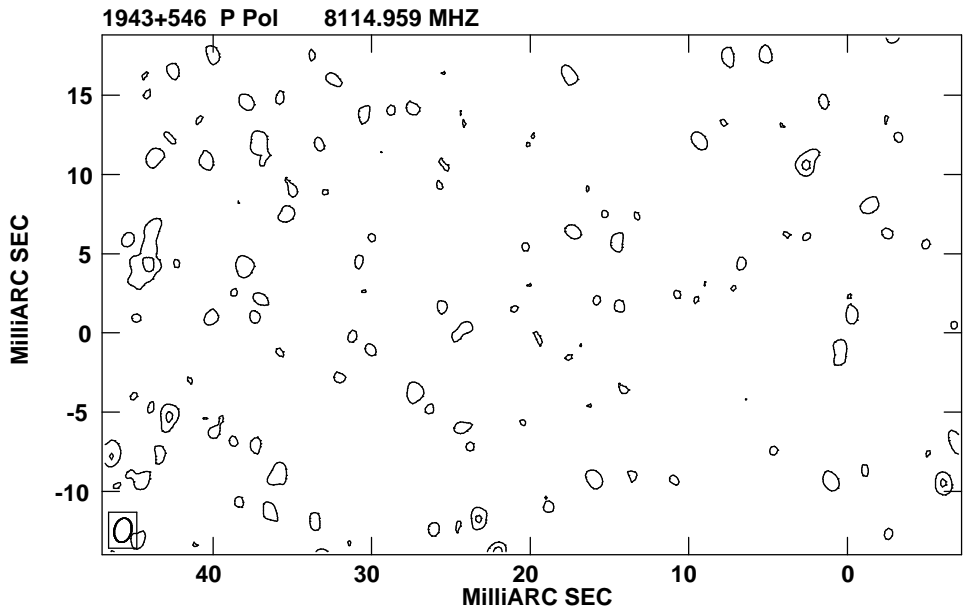}
\caption{Full resolution polarised emission of B1943+546. The first
  contour is 0.3 mJy/beam, and the levels scale up by a factor of
  2. This image has been corrected for the positive bias.}
\label{PPOL}
\end{figure*}

\subsection{Galaxies}
As already mentioned, the integrated linear polarisation in
CSS/GPS/HFP sources identified with galaxies is generally consistent
with being 0.0\% at dm and even at cm wavelengths. Indeed, the absence
of linear polarisation can be used also as a tool to distinguish small
and young radio sources from flat spectrum compact radio sources when 
observed at arcsecond resolution.

One possibility to explain such characteristic refers to a complicated
magnetic field geometry implying a largely reduced integrated
polarisation when averaging over vectors with globally random
distribution. The same result is obtained if an ordered linearly
polarised emission crosses a region of space where there is a
substantial electron density and a randomly oriented magnetic
field, whose characteristic structure happens to be very small, of the
order of a few pc.

CSS or GPS galaxies sources like \object{B0404+768} or
\object{B1943+546} are unpolarised in VLA observations at cm
wavelengths. Their angular size are about 150 and 40 mas
respectively, and then their study requires polarimetric VLBI
observations. They have been observed in a program mainly aiming to
investigate local spectral ageing in their different regions (Murgia
et al. in preparation), but also polarimetry has been carried out. The
images at 8.4 GHz of latter source in total intensity (IPOL) and
polarised emission (PPOL) are presented in Fig. \ref{IPOL} and
Fig. \ref{PPOL}, respectively.  The source structure (Fig. \ref{IPOL})
consists of an unresolved core and two asymmetric lobes, whose radio
emission at this frequency is dominated by the hot-spots and the
beginning of backflow tails. 
Polarisation images turned out to consist of pure noise, indication of
lack of any substantial polarised emission at a level of a fraction of
a mJy. The PPOL image in Fig. \ref{PPOL} does not show any sign of
emission related to the total intensity. Also a tapered image (not
shown) obtained to recover most of the flux density in the radio lobes
does not show any sign of polarised emission. The local fractional
polarisation has then upper limits ranging from 0.2\% in the western 
hot-spot up to a few \% in the more diffuse lobe emission, but it is
likely that the intrinsic linear polarisation at this frequency is
much lower.

A similar conclusion can be drawn for \object{B0404+768}: the highest
frequency with polarisation images available is 5.0 GHz (not shown
here; Morandi et al. in preparation): the peak in the PPOL image is
less than 1 mJy at 5.0 GHz, with upper limits to the linear
polarisation of about 0.4\% in the South-Western hot-spot and,
similarly to \object{B1943+546}, of a few percent in the regions
within the lowest IPOL contour.  

Similar results can be also found in a few GPS galaxies observed at 2
cm with the VLBA by Stanghellini et al. (\cite{cs01}). Thanks to the
relatively short wavelength, some amount (at $\sim$\% level) of linear
polarisation is locally detected on the brightest regions (never in the
cores), but most of the source structure appears unpolarised.

As mentioned above, it is possible to conclude that beam
depolarisation is likely NOT to be the reason of the low linear
polarisation seen in GPS (and smaller CSS) galaxies.
Then, either the screen inhomogeneities are not resolved on scales of a
few pc, or the magnetic field within the radio source is not
ordered (both possibilities may play a role at the same time). 

\subsection{Quasars}

In this section the polarisation properties of a few, well known,
CSS quasars are briefly discussed. In general they have a variety of
behaviours, with no obvious link to other observed quantities like the
projected linear size, turnover frequency and so on. 

Famous examples are \object{3C48}, \object{3C138}, \object{3C147} and
\object{3C286}, also well known to be calibration sources for the
absolute orientation of the electric vector on the sky. Further
examples of well known bright CSS quasars with substantial polarised
emission are \object{3C216} and \object{3C380}.  

A few others possess unusual polarisation properties: L\"udke et
al. \cite{el} show that \object{3C190} does not have polarised
emission at 6 cm in their MERLIN observation, but they quote that the
fractional polarisation has been found as high as 10\% in VLA data at
2 cm (van Breugel et al. \cite{vb92} implying a strong depolarisation
between these two wavelengths. 

\object{3C138} is a well known quasar whose pc-scale radio structure
is know since the early '80s, and whose polarisations properties have
been investigated by Dallacasa et al. \cite{ddb}; Cotton et
al. \cite{wdc97} and \cite{wdc03}. This source is particularly
interesting since it shows a bright jet with polarisation of
$\sim$~10\%, and a core region with a much weaker polarisation
($\sim$3\%) where the true core and another component, likely the first
knot at the beginning of the main jet, are embedded in a diffuse
emission. This knot has some linear polarisation. A monitoring program
carried out with full polarisation to study the motion of this knot
has provided the following evidence: the motion of the knot is at a
marginal level (contrary to previous claims of superluminal motion);
its polarisation properties changed in a way consistent with the
possibility that the inner jet component is seen through limited holes
in a dense Faraday screen. \\
Contrary to all this the polarisation of the main jet, the major
contributor to the total polarised emission, did not show any sign of
variability in both intensity or orientation of the electric vector.

Another quite famous source is \object{3C286}: the interpretation of its
structure is still matter of debate. The milliarcsecond emission is
dominated by a well resolved component, whose structure is similar to
those seen in hot-spots in FR-II radio sources.
The brightest region seen with VLBI is significantly polarised ($\sim
10\%$), and the electric vector does have a configuration resembling
(again) that seen in hot spots.
The polarisation angle is known to be constant at cm and dm wavelengths,
and then the RM is negligible.

\object{3C147} has total intensity morphological characteristics
similar to those seen in \object{3C286}, but the polarised emission do
behave completely different. In fact substantial polarisation is 
detected at wavelengths shorter than 10 cm in the region with the
highest brightness temperature, generally referred to as the core
region. The observed RM (see Rossetti et al. these proceedings; Zhang
et al \cite{zh}) reaches about $-$1650 rad\,m$^{-2}$, that converts to
$\sim$$-$4000 rad\,m$^{-2}$ in the source frame. The polarisation of the
core region is resolved into a number of individual components with RM
in the range  between $\sim$$-$1200 and -2400 rad\,m$^{-2}$.
Once the Faraday rotation is removed, the geometry of the source
magnetic field becomes similar to the one seen in \object{3C286}
reinforcing the similarities between these two sources.

Finally, the it is worth to mention the case of \object{OQ172}. For
some time in the 80's this quasar had the record for the highest
redshift (z=3.53). The radio structure has been studied in detail by
Udomprasert et al. (\cite{udo}) with polarimetric VLBA observations:
some amount of polarised emission arises from the jet base, in which
an extremely high RM is detected and converts up to 40000 rad/m$^2$ in
the source frame. The direction of the intrinsic magnetic field is
aligned with the jet local axis. 

\section{Summary}
The polarisation properties of the CSS/GPS sources are better seen at
cm wavelengths, where depolarisation due to differential Faraday
Rotation within the resolution element is less effective. Linear
polarisation is not detected in GPS and in the smallest CSS galaxies.
Beam depolarisations does not seem to be the reason for this, unless
the Faraday screen has scales smaller than about 1 pc.\\
Some CSS quasars appear outstanding in linearly polarised
emission. Integrated fractional polarisations as high as 10\% are
sometimes observed over a wide range of frequencies.\\
Faraday Rotation is observed in many cases, with RM in the source
frame in the range between  a few hundreds and a few thousands
rad/m$^2$, although there are also cases consistent with zero RM
(e.g. \object{3C286}) or values ad high as 40000 rad/m$^2$
(\object{OQ172}).

\begin{acknowledgements}
Financial support from the Italian MIUR under grant
COFIN-2002-02-8118 is acknowledged.
\end{acknowledgements}

\end{document}